\documentstyle[twocolumn,prc,aps,epsfig,graphicx]{revtex}
\input epsf
\newcommand{\be}{\begin{eqnarray}}
\newcommand{\ee}{\end{eqnarray}}
\newcommand{\bw}{\begin{widetext}}
\newcommand{\ew}{\end{widetext}}
\newcommand{\ds}[1]{
#1{\hskip-2.0mm}/
}

\begin{document}
\twocolumn[\hsize\textwidth\columnwidth\hsize\csname @twocolumnfalse\endcsname

\title {
{Proton Electro-Magnetic Form-Factors in the Instanton 
Liquid Model}}
\author {P. Faccioli, E. V. Shuryak}
\address {Department of Physics and Astronomy, State University of New York,
     Stony Brook, NY 11794
}
\date{\today}
\maketitle
\begin{abstract}
We report the first non-perturbative
calculation of proton electro-magnetic 
form-factors in the
Random Instanton Liquid Model (RILM) and in the Interacting Instanton Liquid 
Model (IILM). By 
calculating the ratio of appropriate three-point 
 to  two-point functions, we divide out the  coupling constants
and compare our results
directly to some integral of the form-factors.  
Using various parametrizations of the electric
form-factor $G_E(Q^2)$ at large $Q^2>3.5\, GeV^2$,
 where it is not yet measured, we
compare
those with expected theoretical dependence.
We find from this comparison that some distributions of charge
are clearly excluded (e.g. the same as of the
magnetic moment, $\mu G_E/G_M=1$, as well
as the opposite scenario in which $\mu G_E/G_M$ rapidly approach
zero), restricting possible behavior of the form-factor to rather
narrow band.
 Furthermore, we found that our
calculation of the nucleon form-factors is dominated by a configuration
in which two out of three quarks interact with a
  $single$-instanton, in spite of the fact that the evaluated
 three-point function 
has rather large distances ( $\sim 1.2 fm$ ) between points. We also
estimate the size of the scalar di-quark and found it to be very small,
comparable to the  typical instanton size. 

\end{abstract}
\hspace{12 cm} PAC number(s): 11.15tk
\vspace{0.1in}
]




\newpage
\section{Introduction}
\label{intro}

The non-perturbative sector of QCD is not yet theoretically understood,
and any direct comparison between high accuracy
experimental data and theoretical
predictions is very valuable. 
The electro-magnetic form-factors of hadrons are fundamental
observables,  depending on one scale $Q^2$, which are related to
 the probability of an hadron to absorb a photon and yet 
remain the same hadron.
Apart of telling us much about hadronic sizes and structure, and about
 the dynamics responsible for the ricombination of partons in the final state,
form-factors  are also supposed to teach us where the transition from the
non-perturbative to perturbative regime takes place.

Renewed interest to these issues is
triggered by the new precision
measurements pursued  at the
Jefferson Laboratory. Specifically, their
 new measurements  (using a polarization-transfer method) 
have yielded more accurate  proton form-factors, 
 showing for the first time  that distributions of charge and magnetic moment
 are not identical, $\mu\,G_E/G_M \ne 1$, \cite{jlab}.  
The ratio $\mu\,G_E(Q^2)/G_M(Q^2)$ was  measured up to 
$Q^2=3.47 \,GeV^2$ and new data up to $Q^2\simeq 5.6 \,GeV^2$ should be
available soon.
These result are plotted at fig. \ref{GEGM} below.

Theoretical arguments about this issues are rather controversial.
Apparent agreement between existing data and
 pQCD power counting \cite{counting1},\cite{counting2} ($F_\pi(Q^2)\sim 1/Q^2,G_N\sim
1/Q^4$) have lead some to think that perturbative regime is reached (for 
a review of pQCD analysis of hadronic form-factors see, e.g. 
\cite{sterman}).
However, at least 
in the case of the pion form-factor we do know that the observed value 
of $Q^2\,F_\pi(Q^2)$ is different from its asymptotics at infinite $Q^2$,
roughly by factor two,  and thus some adjustment of its value 
at higher $Q^2$ than measured so far should take place.

There are also numerous arguments which show
that the origin of the form-factors at $Q^2< 2 \,GeV^2$ is
non-perturbative.  
Historically, the first calculations of this kind for the proton form-factors 
were based on the Operator Product Expansion and QCD
sum-rule analysis, supplemented by the assumption of vacuum
dominance \cite{shifman80,ioffe,radyushkin}. In this approach all
information about the non-perturbative  properties of the QCD 
vacuum  are parametrized  via few quark and gluon condensates.
However, this approach is only valid at small enough distances, and
 in order to extract physical observables 
 one is then forced to consider complete experimental 
 spectral densities
 which includes contributions from multiple 
excited states, not just pion or nucleon alone.
To overcome this difficulty, a simple ``pole-plus-continuum'' (PPC) 
model for the 
spectral representation is often adopted.
Unfortunately, that introduces 
significant  model dependence in the analysis and reduces its
predictive power. In practice,  this approach is not always  
successful: for example 
the sum-rule calculation of the proton's two-point
function involving a 
PPC ansatz agrees
with the experimental data, only  if a particular Ioffe current is used
\cite{DoschI}.
\begin{figure}
\vskip 0.2in
\includegraphics[width=2.5 inc]{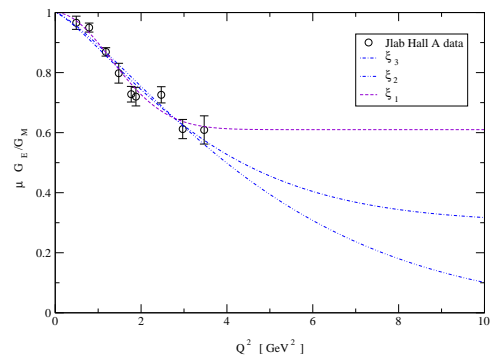}
\vskip 0.2in
\caption[]{
\label{GEGM}
JLAB results for $\mu\,G_E/G_M$ \cite{jlab}
as compared to the fits (\ref{fit}) with (\ref{ansatz}).}
\end{figure}

\begin{figure}
\vskip 0.2in
 \includegraphics[width=2.5 inc]{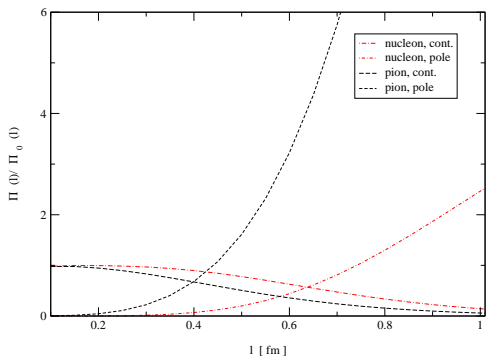}
\vskip 0.2in
 \caption[]{
\label{pole_cont}
Pole and continuum contributions for the pion \cite{shuryakcorr}
 (dashed, $s_0\,= \,1.6 \,GeV^2$) 
and nucleon (dot-dashed, $s_0\,= \,2.5 \,GeV^2$) two-point correlation functions.
}
 \end{figure}

In general, hadronic data usually tell us about long-distance
behavior,
while field theoretic calculations have better predictive power in the
opposite limit of small distances. One issue any application
of such method has to face
is  the {\em existence} and size of the {\em ``working window''} of distances
in which a quantitative comparison between the two can be made.
  To be successful, a theoretical approach
  has to be able to  predict the correlation functions at
relatively large ($\sim 1 fm$)  distances:
then the unknown contribution from all excited states is automatically 
suppressed and only
the pole corresponding to the lowest lying particle survives.
Two illustrative examples of relative strength of both components, for 
two-point 
isovector pseudoscalar current correlator (``pion'') and the nucleon
(with current to be specified below) are shown in Fig.\ref{pole_cont}
as a function of (Euclidean) distance $l$ between the currents.
Here and below, all correlators are normalized to their free quark
value (so unit value at small $l$ means just free quark motion,
a simple consequence of asymptotic freedom).
Frome those examples one can see, that the pole dominance is rapidly
reached at some (channel-dependent) value of $l$,
 about $0.8 \,fm$ for the nucleon.

The most direct way of calculating {\it ab initio} such correlation functions
is numerical lattice-based approach. However,
 quite successful results were also obtained from much simpler
instanton-based models, see review \cite{shuryakrev}
and also \cite{SS_01} for recent example of quantitative predictions
of the model, compared to the $\tau$ lepton decay data. 
Closest to the present work is
the successful calculation of the pion form-factor 
\cite{forkel,blotz}
using the  same instanton-based  model as used below. 
One important aspect of \cite{blotz} is the direct relation found
there between the pion and instanton sizes. 

  In the present work we try to answer the following questions.
How to organize the calculation, what are the best combination
of three and two-point
correlators to be used?
At what distances those should be calculated, in order to exclude
contributions of the nucleon excited states?
 How do the results obtained correlate with
experimental form-factors? Are they sensitive to the high-$Q^2$
part which is not yet measured, and if so, can we make some
predictions? Can the proton form-factors be described in terms of
 configurations of three valence quarks bound by instanton forces?
Are these dominated by one single instanton, at least at some scale?
If so, do all three quarks interact strongly with it, or only two, as
arguments based on Pauli principle for zero modes would suggest?
And finally, how does this approach
correlate with phenomenological models of the nucleon, such as
quark-diquark model? 

The paper is organized as follows.
In section \ref{larged}, we will present our  results for the 
nucleon two-point and electro-magnetic three-point function, obtained 
in the Random Instanton Liquid Model (RILM),  in the Interacting
Instanton Liquid Model (IILM) and in the Single Instanton Approximation (SIA).
Is section \ref{pheno} we show how such correlation functions 
can be evaluated from the phenomenological knowledge of 
proton mass and form-factors and we compare them  
with our theoretical predictions.
In section \ref{di-quark} we discuss the physical implications of our
calculations, in particular in relation to different hadronic models of the
nucleon. 
Results and conclusions are summarized in section \ref{conclusions}.


\section{Calculation of the Correlation Functions  from 
the Instanton Liquid Models.}
\label{larged}

Our theoretical investigation is based on the analysis of two and three-point
Euclidean correlation functions, in coordinate space. 
In particular we consider the electro-magnetic three-point function: 
\begin{equation}	
\label{threepoint}
\Pi_{sc \, 4}(x,y)=<0|\, tr\,[ \, \eta_{sc}(x/2) \, J^{em}_4(y)\bar{\eta}_{sc}(-x/2) \, \gamma_4 \,]\,|0>,
\end{equation}
where $J^{em}_4(y)$ is the fourth component of the electro-magnetic current,
\begin{equation}
J^{em}_\mu(y):= \frac{2}{3}\, \bar{u}(y)\,\gamma_\mu\, u(y) -  \frac{1}{3}\,
 \bar{d}(y)\,\gamma_\mu d(y).
\end{equation} 
$\eta_{sc}(x)$ denotes a particular combination of Ioffe currents, that 
we refer to  as the scalar current:
\begin{eqnarray}
\eta_{sc}(x)  := \epsilon_{a b c} [ u^a(x) \,C \gamma_5\,  d^b(x) ]\, u^c(x), 
\end{eqnarray} 
where $a, b, c$ are color indices.
In principle, one can choose any linear combination of Ioffe 
currents, which  excite states with the
same addictive quantum numbers of the proton 
(as well as states with opposite parity).
Our choice  is theoretically motivated by the fact that  
$\eta_{sc}^c(x)$ contains explicitly a di-quark term, which was shown 
to receive maximal coupling to instanton's zero modes \cite{shuryakrev}. 

The nucleon coupling to such current is defined as
\begin{equation}
\label{lambdas}
<0| \eta_{sc}(x)| p,s> = \Lambda_{sc}  u(p,s) e^{-i q\, x}, 
\end{equation}
where $u(p,s)$ is a Dirac spinor and $s$ is the spin index.

Let us specialize now a particular choice \footnote{Notice that, 
in such configuration, the vertices
 $x/2$, $-x/2$ and  $y$ do not lie 
on the same line. The convenience of  such choice 
will be discussed  in the next section.}
 for the three-point function 
(\ref{threepoint}),
in which:
\be
\label{configuration}
y\cdot x =0, \qquad {\it and}\nonumber \\
l:= |y|=|x|.
\ee

We shall consider also 
the following two-point Euclidean proton correlation function
\begin{equation}	
\label{twopoint}
\Pi_{sc}(x)=<0|\, Tr\,[ \, \eta_{sc}(x/2) \, \bar{\eta}_{sc}(-x/2) \, 
\gamma_4 \,]\,|0>,
\end{equation}
in which
the propagation can been chosen along the Euclidean time axis.
Due to the presence of  $\gamma_4$,  this correlator
receives  contribution  from two quark zero-modes propagators 
in the field of an instanton
\footnote{This can be verified by simply counting the quark chirality flips.}.
Notice that, in a theory with two massless flavors, this is the maximal 
number allowed by Pauli principle.

Since  the same factor $\Lambda_{sc}^2$, which can not be determined from
 experiments, appears both in the expression for the proton two-point function
(\ref{twopoint}), and in three-point function (\ref{threepoint}), it is 
natural to consider the ratio:
\be
\label{Gamma}
\Gamma(l):= \frac{\Pi_{sc\,4}(l,l)}{\Pi_{sc}(l)},
\ee
in which such factor is divided out.  
In the next section we show that  $\Gamma(l)$ 
for $l\gtrsim 0.7 fm$  depends only on the
the proton's mass and form-factors.
At this scale, $\Gamma(l)$ is dominated by non-perturbative effects 
and therefore  represents a suitable tool to probe model descriptions
of the non-perturbative strong dynamics.

In instanton models, one assumes that the 
Euclidean  QCD partition function is saturated by the configurations 
of an interacting ensemble of pseudo-particles 
(instantons and anti-instantons) 
which, in ordinary Minkowski space, are associated to the tunneling between 
degenerate classical vacua.

Each  configuration of the ensemble is determined by a set of 
collective coordinates, specifying the
 the position, the size and the orientation in color space
of each instanton.

The parameters of the model are the average instanton size,
 $\bar{\rho}\simeq 1/3\, fm$ and the average instanton density, 
$\bar{n}\simeq 1 fm^{-4}$.
These values have been determined from
 several phenomenological studies \cite{shuryak82},\cite{shuryakrev} 
as well as from lattice simulations \cite{chu94}.

In the Random Instanton Liquid Model (RILM), one assumes that all instantons
have the same size and that the distribution of all remaining collective 
coordinates is completely random.
In the Interacting Instanton Liquid Model (IILM) the distribution is
calculated from numerical simulation of the partition function, 
describing interaction between instantons. The most important and
complicated part of this interaction is fermion exchanges between them,
 described by the fermionic determinant in the partition function.

Both the RILM and IILM have been shown to 
have nucleon as a bound state, and give
 reasonable values for the 
its mass \cite{shuryakrev}. Furthermore, these models 
provide very different nucleon
(octet) and delta (decuplet) correlators, explaining the 
splitting between the nucleon 
and the delta in terms of zero mode forces \cite{baryon2pt}, 
\cite{baryon2ptint}.
Another important result of these works is the prediction of the
nucleon coupling constants to Ioffe current: it is proportional to the
probability to find all three quarks at the same point, and is needed for 
such applications as evaluation of proton decays.

  As the first step toward understanding of the function  $\Gamma (l)$
  we have evaluated  it
analytically 
in the Single Instanton Approximation (SIA), in which only the closest instanton is taken into
 account explicitly \cite{shuryak82}, \cite{SIA}. 
In such an approach, the correlation functions
are evaluated from the 
the quark propagator $S_{I(A)}(x,y)$ in the instanton 
(anti-instanton) background.
This is known exactly to be the sum of two terms:
a zero-mode part, $S_{zm}(x,y)$ and a non-zero mode part,
 $S_{nzm}(x,y)$.
In most applications, $S_{nzm}(x,y)$ can be approximated with the free 
propagator, $S_0(x,y)$, while the quark zero-mode propagator reads:
\be
\label{s}
S_{zm}(x,y) =\frac{\psi_0(x)\,\psi^\dagger_0(y)}{-i\,m},
\ee
where $\psi_0(x)$ represents an eigenmode of the Dirac operator
\footnote{The explicit expression for $\psi_0(x)$ is, of course, 
gauge dependent. The ILM is formulated in the singular gauge, where the 
topological charge is localized around each instanton.}
 with vanishing 
eigenvalue (quark zero-mode),
\be
i\ds{D}\,\psi_0(x) &=& 0,
\ee
\be
\label{zmodes}
\psi_{0\,\,a\,\nu}(x;z) = \frac{\rho}{\pi}
\frac{1}{((x-z)^2+\rho^2)^{3/2}}&\cdot&\nonumber\\\cdot
\left[
\frac{1 - \gamma_5}{2}\, \frac{\ds{x}-\ds{z}}{\sqrt{(x-z)^2}}
\right]_{\alpha\,\beta}\cdot
U_{a\,b}\,\epsilon_{\beta\,b},
\ee
(here $z$ denotes the instanton position, 
$\alpha, \beta= 1,..., 
4$ are spinor indices and $U_{a b}$ represents a 
general group element).

In \cite{SIA} we showed that, if the correlation function receives contribution
from at least two quark zero-mode propagators,  
one can incorporate the effects of all
 other instantons in an effective mass term that 
replacing the current mass in (\ref{s}). 
More specifically, correlation functions receiving contribution 
from two zero-mode propagators
will be proportional to inverse powers of the current quark mass,
 $m^2$, which will be replaced by an
 effective parameter  $m_2^2$.
Such parameter depends on the ensemble and has been determined to be $m_2^2 \simeq (65\, MeV)^2$ for the RILM and $m_2^2 \simeq (105\, MeV)^2$ for the IILM.
The resulting expressions for $\Pi_{sc}(l)$ and $\Pi_{4\,sc}(l,l)$ are
\footnote{Our theoretical calculations for $\Pi_{4\,sc}(l,l)$ have been
obtained retaining  only connected diagrams. 
This is motivated by the fact that all one-instanton contributions
to the disconnected diagrams for the electro-magnetic three-point function
are mqzero.}:
\be
\Pi_{sc}(l)=\frac{15}{\pi^6\,l^9} + \frac{12\,\bar{n}\,\rho^4}{l^3\,\pi^4\,m_2^2}
\,\int \,d^4 z\cdot\nonumber\\ 
\cdot \frac{1}{
[\vec{z}^2+(z_4-l)^2+\rho^2]^3\,
[\vec{z}^2+z_4^2 +\rho^2]^3}, 
\ee
and
\begin{eqnarray*}
\Pi_{4\, sc}(l,l) =
\frac{96}{25\, \pi^8\,l^{12}}\,+ 
\frac{4\,\bar{n}\,\rho^4}{l^6\,15\,\pi^8\,m_2^2}
\int d^4 z 
\hspace{1 cm}
\nonumber\\
\cdot \frac{1}
{\left[ 
(l/2 + z_4)^2 + \vec{z}^2 +\rho^2
\right]^{3/2}\,
\left[ 
(l/2 - z_4)^2 + \vec{z}^2 +\rho^2
\right]^{3/2}}
\nonumber \\
\cdot
\left\{ 
\frac{8}{\left[ 
(l/2 + z_4)^2 + \vec{z}^2 +\rho^2
\right]^{3/2}\,
\left[ 
(l/2 - z_4)^2 + \vec{z}^2 +\rho^2
\right]^{3/2}} \, + \right.
\nonumber\\ 
\left. 
\frac{7/5}{\left[ 
 (l - z_2)^2 + Z^2 + \rho^2
\right]^{3/2}
\left[
 (l - z_2)^2 + Z^2 
\right]^{1/2}}
\right. \nonumber \\
\left. 
\cdot\left(\frac{2\,l^2 + 5\,l\,z_4+\vec{z}^2+2\,z_4^2-4\,l\,z_2 }
{\left[ 
(l/2 - z_4)^2 + \vec{z}^2 +\rho^2
\right]^{3/2}\,\left[ 
(l/2 + z_4)^2 + \vec{z}^2 
\right]^{1/2}}+
\right.\right.\nonumber\\
\left.\left.
\frac{2\,l^2 - 5\,l\,z_4+\vec{z}^2+2\,z_4^2-4\,l\,z_2 }
{\left[
(l/2 + z_4)^2 + \vec{z}^2 +\rho^2
\right]^{3/2}\,\left[ 
(l/2 - z_4)^2 + \vec{z}^2 
\right]^{1/2}}
\right)
\right\}, 
\end{eqnarray*}
\be
\mbox{}
\ee
where 
 $\vec{z}^2:= z_1^2 + z_2^2 + z_3^2$ and $Z^2:= z_2^2 + z_3^2 +z_4^2$.

Note, that
contribution from all other instantons enters only via
an effective mass parameter.
Although the values of these effective masses in RILM and
IILM
are quite different. Fortunately, one finds  that 
two and three-point functions have the same power of this parameter, and
thus it
actually drops out in our ratio $\Gamma(l)$ as soon as the
 perturbative (free) contribution becomes negligible.
 
\begin{figure}
\vskip 0.2in
 \includegraphics[width=2.5 inc]{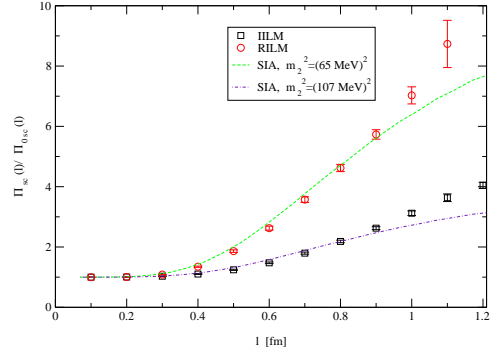} 
\vskip 0.2in
 \caption[]{
\label{siafig2p}
Proton two-point function, $\Pi_{sc}(l)$ evaluated in 
the ILM and in the SIA.
The circles (squares) correspond to RILM (IILM) results.
Dashed (dot-dashed) lines correspond to SIA results using the mass parameters 
evaluated in \cite{SIA}: 
$m_2^2= (65 \,MeV)^2$ ( $ m_2^2= (107 \,MeV)^2 $ ).}
 \end{figure}

\begin{figure}
\vskip 0.2in
\includegraphics[width=2.5 inc]{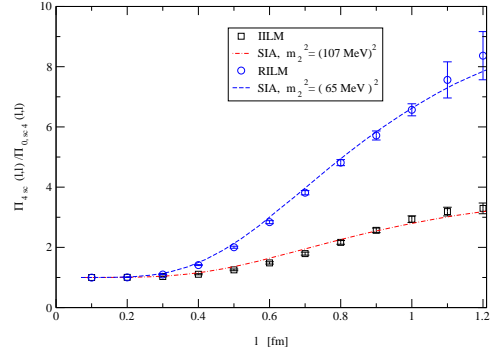}
\vskip 0.2in
\caption[]
{
\label{siafig3p}
Proton three-point function, $\Pi_{4\, sc}(l)$ evaluated in the 
ILM and in the SIA.
The circles (squares) correspond to RILM (IILM) results.
Dashed (dot-dashed) lines correspond to SIA results using the mass parameters 
evaluated in \cite{SIA}: $m_2^2= (65 \,MeV)^2$ ( $ m_2^2= (107 \,MeV)^2 $ ).}
 \end{figure}

The SIA is supposed to break down when quarks 
are propagating for distances larger than the typical separation between
two neighbor instantons ($\sim 1 fm$).
Nevertheless, in \cite{SIA} we showed that the
 SIA calculations for hadronic 
two-point functions agree with the 
corresponding ILM results up to quite large distances 
($\sim 0.7\div \sim 1 fm$, depending on the correlation function) 
so there may  be a window for such an approach in the present calculation.
In figs. \ref{siafig2p} and \ref{siafig3p} we have plotted the
ratios $\Pi_{sc}(l)/\Pi_{sc\, free}(l)$ 
and $\Pi_{4\, sc}(l,l)/\Pi_{4\, sc\, free}(l,l)$ obtained in the SIA
and from numerical simulations in the RILM and IILM
 \footnote{In producing these plots we have
used the effective mass parameters defined and evaluated in \cite{SIA}.}.
It is quite surprising to notice that the SIA results for three-point
 functions reproduce the corresponding ILM result all the way up to $1.2 fm$.
For larger distances  statistical errors on ILM results make the comparison
somewhat meaningless.

\begin{figure}
\vskip 0.2in
\includegraphics[width=3.1inc, clip=]{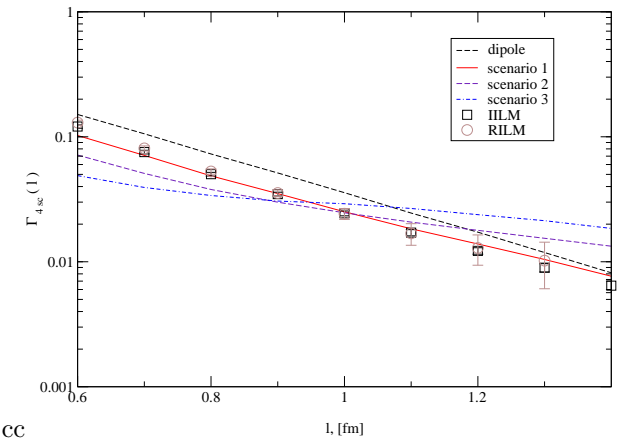}
\vskip 0.2in
 \caption[]{\label{Gammafig}
Large distance behavior of $\Gamma(l)$.
The lines represent the contribution of the nucleon pole, for the
different scenarios given by (\ref{fit}) and (\ref{ansatz}).
In particular, the solid line corresponds to the scenario 1, 
 where the ratio $\mu\,G_E/G_M$ stabilizes at around 
$0.6$ at high $Q^2$, while the large-dashed line was obtained using
the  dipole formula for \emph{both} $G_M$ and $G_E$.. 
The circles  (squares) represent RILM (IILM) predictions. }
 \end{figure}

Now we make the second step, and 
 numerically evaluate the function $\Gamma (l)$ from
a random and an interacting instanton ensembles.
In both cases we used propagators which include complete
re-diagonalization
in the zero-mode subspace. 
We have used $256$ instantons and anti-instantons
in a  $(4 fm)^4$ periodic box.
The results are presented in fig. \ref{Gammafig}:
 one can see that, although the IILM and the RILM 
give  different two-point and three-point
correlation functions (see figs. \ref{siafig2p} and \ref{siafig3p}), their 
 ratio $\Gamma (l)$ is essentially the same.

\section{Relating the Euclidean correlation functions with the
experimental data}
\label{pheno}

In a field theoretic Euclidean framework (such as lattice QCD or
instanton-based models)
a very important step is to encode available
 information about masses, form-factors, wave 
functions and other properties 
of particular hadrons into appropriate Green functions.

The spectral representation of  correlators can be traditionally written 
assuming a ``pole-plus-continuum'' ansatz for the spectral density:
\begin{equation}
\label{2ptspectral}
\Pi_{sc}(x) = -(\Lambda_{sc})^2 D'(M_N,x)- \frac{1}{64\, \pi^4}\, 
\int_{s_0}^{\infty} \, ds\, s^2 \,D'(\sqrt{s},x),
\end{equation} 
Here the first term represent the nucleon contribution:
 $M_N$ is the nucleon mass and  $D(m,x)$ is the Euclidean 
scalar propagator of a particle of mass $m$.

The second term represents the contribution of the excited states
with nucleon quantum numbers (including those with opposite parity).
This particular form of ``continuum contribution''
represents parton-hadron duality, it corresponds
to the discontinuity of the perturbative (free)
quark loop diagram contributing to the same correlation function.
The new parameter $s_0$ identifies 
the threshold for the onset of the duality regime: at this energy
transition from the non-perturbative to
 the perturbative regime takes place. The QCD sum rules practitioners
obtain the value for this parameter  together with others
(masses and coupling constants), from comparing  theoretical and
experimental dependence of the correlator.
However, the results still depend on this specific assumption about
the contribution of the excited states.

We have chosen not to deal with this uncertainty, calculating all
correlation functions 
at sufficiently large distances $x\gtrsim 1 fm$,
where the continuum contribution is negligible 
compared to nucleon contribution.
(see fig. \ref{pole_cont}).

The contribution to (\ref{threepoint}) due to the nucleon pole can be
 evaluated in terms of the Fourier transforms of the Dirac and Pauli 
form-factors (see appendix \ref{deriv} for the detailed derivation):
\be
\label{nucleonpole}
\Pi_{sc\,4} (x,y) \stackrel{|x|,|y| \gg 1}{\rightarrow} \Lambda_{sc}^2 \int\,d^4 z\, Tr\,
\left[ 
S_{M_N}\left(\frac{x}{2}, y+z \, \right)
\right.\nonumber\\
\left.\left[\gamma_4\,F_1(z) + \frac{i\sigma_{4\,\nu}} {2\,M_N} \,F_2^\nu (z)\right] 
S_{M_N}\left(y+z,-\frac{x}{2} \, \right)\,\gamma_4\,
\right] \nonumber,
\\
\ee
where $S_{M_N}(x,x')$ denotes the nucleon propagator from $x'$ to $x$ and
\be
F_2^\nu(z) := i\,\partial^\nu\, F_2(z)=\frac{i\,z^\nu}{|z|}F_2'(z).
\ee

\begin{figure}
\vskip 0.2in
\includegraphics[width=2.5 inc]{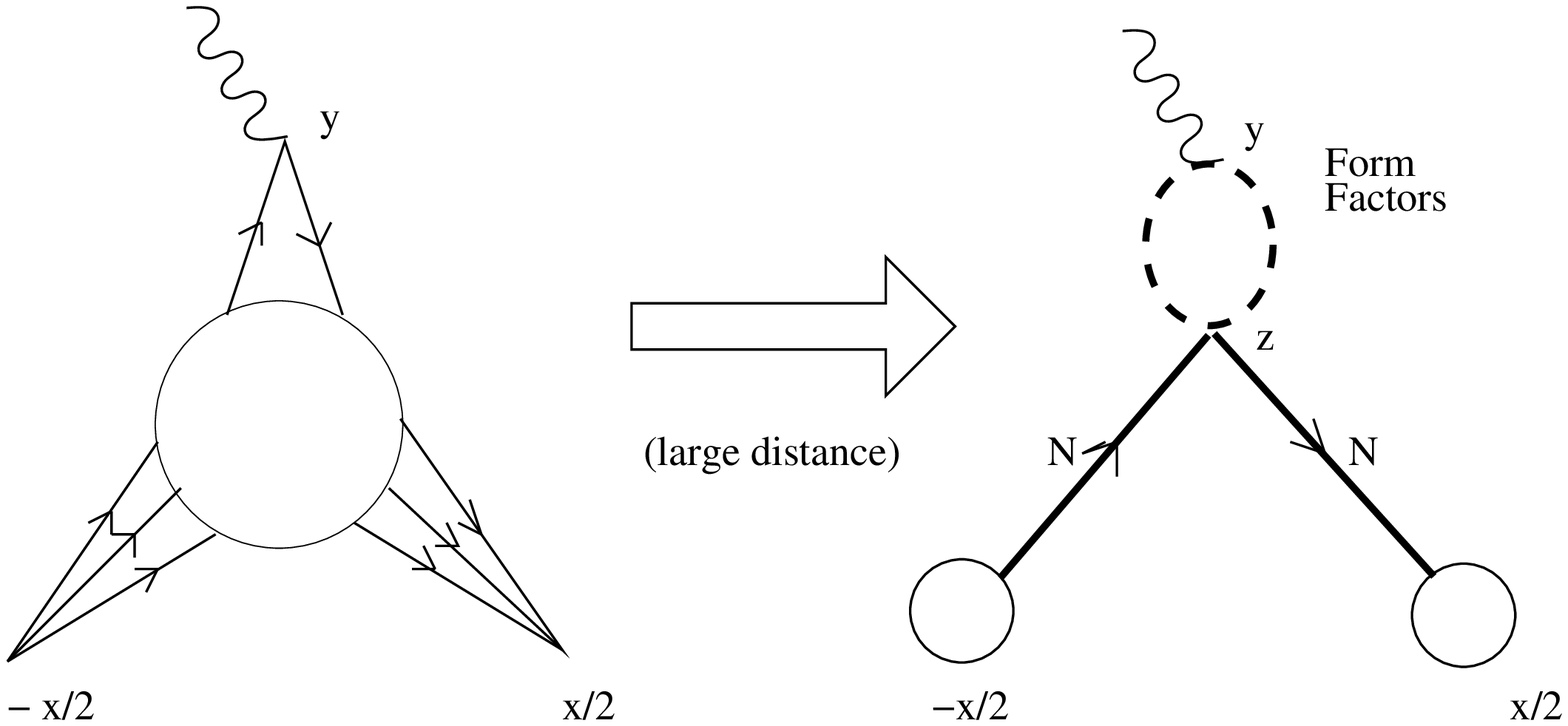}
\vskip 0.2in
 \caption[]{
\label{geometry}
In the large distance limit,
the three-point function $\Pi_{Sc 4}$  is dominated
by the nucleon pole.}
 \end{figure}

Eq. (\ref{nucleonpole}) has a simple physical interpretation 
(see fig. \ref{geometry}): at large 
distances, the three-point function (\ref{threepoint}) represents the 
(Euclidean) amplitude
for a finite-sized nucleon to be created at $-x/2$, 
to absorb a photon at $y$ and 
be annihilated at $x/2$.
Notice that the electro-magnetic vertex depends on the Fourier transforms of
the Pauli and Dirac form-factors 
and therefore contains  information about the shape of the proton.

In order to evaluate $\Pi_{sc\, 4}(x,y)$, we need an expression 
for $F_1(z)$ and  $F_2(z)$. 
This can be obtained by Fourier transforming the analytic 
continuation to Euclidean space of the fits of the
 experimental form-factors.
Typically, elastic scattering experiments allow to obtain the
 Sachs electric and magnetic form-factors, which are related to the Pauli and 
Dirac form-factors by:
\be
\label{Ge}
G_E(Q^2) &=& F_1(Q^2) - \tau \, F_2(Q^2)\\
\label{Gm}
G_M(Q^2) &=& F_1(Q^2) +  F_2(Q^2),
\ee
where $\tau := Q^2/4\,M^2$ and $Q^2=-q^2$.

At low momenta, $Q^2 \lesssim 1 \, GeV^2$,  
$G_E(Q^2)$ and $G_M(Q^2)$  can be extracted
from usual Rosenbluth separation method and error-bars are 
 of the order of a few percent. 
In this regime, the data for both form-factors are
well fitted  by the so-called dipole formula, 
\be
\label{G_Mdip}
G_M(Q^2)/\mu \simeq G_E(Q^2) \simeq  G_D(Q^2), \qquad Q^2\lesssim 1 \,GeV^2
\ee
where
\be
\label{dipole}
G_D(Q^2)=\frac{1}{(1+\frac{Q^2}{0.71})^2},
\ee
and $\mu$ is the proton magnetic moment.

As the momentum increases, the cross section becomes
 dominated by $G_M$ making the extraction of $G_E$ more and more difficult.
 As a result, error-bars on $ G_E$ become very large already at 
$Q^2\sim 2 \,GeV^2$ while error-bars on $G_M$ remain small up to 
$Q^2\simeq 30 \,GeV^2$ \cite{GM}.

\begin{figure}
\vskip 0.2in
\includegraphics[width=2.5 inc]{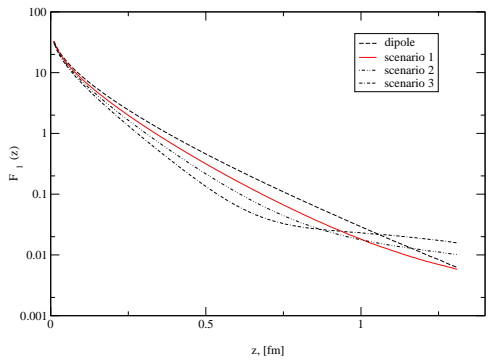}
\vskip 0.2in
 \caption[]{
 \label{f1}
Fits of the Pauli form-factor ($F_1$) in coordinate 
representation obtained using the parametrization (\ref{ansatz})}.
 \end{figure}

\begin{figure}
\vskip 0.2in
\includegraphics[width=2.5 inc]{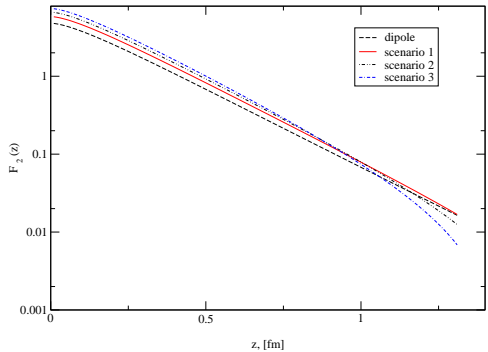}
\vskip 0.2in
 \caption[]{
 \label{f2}
Fits of the Dirac form-factor ($F_2$) in coordinate 
representation obtained using the parametrization (\ref{ansatz})}.
 \end{figure}

Using this experimental information we are now able to complete our 
phenomenological estimate of the relevant three-point function.
The coordinate representation of $G_M$ has been obtained by Fourier
 transforming the dipole fit (\ref{G_Mdip}), which is very accurate at the 
momentum scale we are interested in.
The fit for $G_E$ can be obtained as follows.
One can assume the following ansatz:
\be
\label{fit}
G_E(Q^2)= G_D(Q^2)\, \xi(Q^2),
\ee
where the function $\xi(Q^2)$, which accounts from the deviation from dipole
shape, can be obtained from the available data for $\mu\,G_E/G_M$.
These can be well reproduced by the following set
of exponential fits which, however, have different trends 
in the kinematical region 
$Q^2>3.5\,GeV^2$ ( see fig. \ref{GEGM} ):

\be
\label{ansatz}
\xi_{d}(Q^2) &=& 1. \qquad \emph{(dipole parametrization)}\nonumber\\
\xi_1(Q^2) &=&  0.39\, e^{-\frac{(Q^2)^2}{3.12}} + 0.61.\nonumber\\
\xi_2(Q^2) &=&  0.70\, e^{-\frac{(Q^2)^{1.3}}{5.39}} + 0.30\nonumber\\
\xi_3(Q^2) &=&  e^{-\frac{(Q^2)^{1.3}}{8.71}}
\ee
The corresponding functions $F_1(z)$ and $F_2(z)$ are plotted in fig.
\ref{f1} and 
fig. \ref{f2}.

The results of our phenomenological calculation for 
$\Gamma(l)$, using the parametrization (\ref{ansatz}) for 
$\xi(Q^2)$ are reported in fig. \ref{Gammafig}.
It is important to recall that our phenomenological calculation
of $\Gamma (l)$ is valid only at large distances, where the proton 
pole is dominant.
We therefore need
to estimate how large $l$ needs to be for this assumption to hold.
From the analysis of the two-point function, we have seen that the continuum
contribution is almost completely suppressed for $l\gtrsim 1\, fm$.
In the electro-magnetic three-point function, however, one of the quarks is
struck in the point $y$, acquiring and undetermined amount of momentum.
As a result, at such vertex the nucleon pole  
gets mixed with all other resonances and one needs to wait for another 
$\sim 1\, fm$, in order for the pole to decouple again.
From this rough estimate, one expects that (\ref{threepoint}) 
can be identified with (\ref{nucleonpole}) if
 the three quarks have propagated for about $ 2 fm$.
In the ILM and on the lattice, the vacuum is usually simulated 
in a four-dimensional periodic box, whose maximal size is determined 
by the machine's precision and by computational time constraints.
It is therefore important to make a convenient choice of  $x$ and $y$
 in (\ref{threepoint}).
In particular, we need to consider three-point functions 
that are large enough for
the ground-state pole to decouple but  can be fit into
a relatively small box. 
At this purpose, it is generally a good idea  to let quarks propagate
 along the diagonals.
With the choice (\ref{configuration}) one has that the 
quarks travel for the needed $\sim 2 fm $ already for $l\sim 0.7 fm$.

\begin{figure}
\vskip 0.2in
\includegraphics[width=2.5 inc]{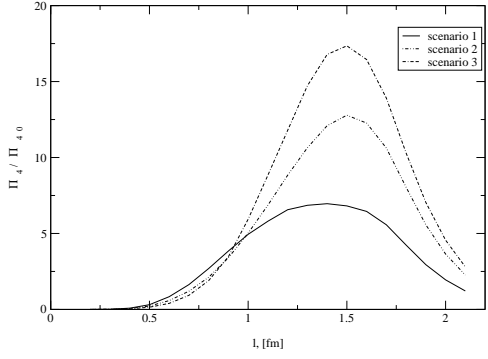}
\vskip 0.2in
\caption[]{
\label{3scenari}
Proton pole contribution to the Electro-magnetic three-point 
function $\Pi_{4\, sc}$, obtained from three different fits for
 $\mu\,G_E/G_M$ (eq. \ref{ansatz})
which differ only for $Q >3.5\,GeV^2$.}
\end{figure}

At this point, we may ask to what momentum scale is our analysis in
configuration space sensitive to. 
From uncertainty principle arguments, one would naively 
expect that correlation functions 
of size $l$ should be essentially sensitive to the physics
at the scale $Q\sim 1/l$. 
Along this line, one is then lead to argue that our large sized 
($l> 0.7\, fm$) 
correlation functions can only be used to study to the small momentum 
part of the form-factors.
This is certainly true asymptotically: for example,
in the (numerically infeasible)
limit of extremely large $l$, the three-point function should depend only on 
the proton electric charge and magnetic moment and, therefore, all scenarios
 considered in (\ref{ansatz}) must give the same answer.
However, at intermediate distances the situation is more complicated.
In fact, eq. (\ref{nucleonpole}) shows that $\Pi_{sc\,4}(l)$, is related 
through a  non-trivial integral to the Fourier transform 
of the form-factors (eq. \ref{nucleonpole}).
Such integration mixes the different modes in such a way that 
three-point functions 
with $l>0.7\, fm$ are still quite sensitive to the shape of form-factors at
high $Q^2$.
This is already seen  in fig. \ref{3scenari}, where $\Pi_{4\, sc}(l,l)$,
 obtained form fits of the form-factors 
that differ only in the high momentum range ( $Q^2> \,3.5 \,GeV^2$)
 are compared.
Clearly,  such green functions at, say,  $l\sim 1.5\, fm$ are quite
 different and larger distances are needed for the three curves to converge.

\begin{figure}
\vskip 0.2in
\includegraphics[width=2.5 inc]{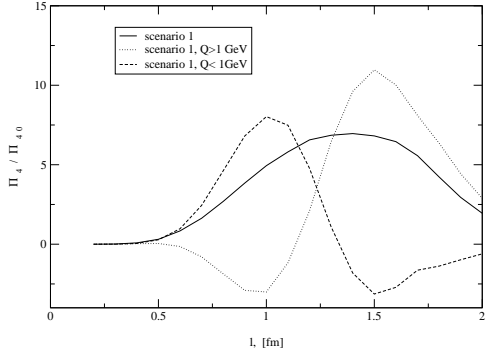}
\vskip 0.2in
\caption[]{
\label{highlowq}
Contribution of different momenta to the proton  electro-magnetic three-point 
function $\Pi_{4\, sc}$ .
The solid line represents the correlators obtained in scenario 1, 
integrating over all momenta, the dotted (dashed) line is obtained 
keeping only the 
$Q> \,1\,GeV$ ($Q< \, 1 \, GeV^2$) components of the form-factors.}
\end{figure}

It is interesting to  further analyze the contribution of different momentum 
components to our correlator.
At this purpose,  in fig. \ref{highlowq} 
we have plotted the three-point function corresponding to 
scenario 1, and we compare it with two more  curves, that one obtains
from the same parametrization, but keeping
only the contribution from $Q< 1\, GeV$ and $Q>1 \,GeV$
to the form-factors $F_1(z)$ and $F_2(z)$ (see eq. \ref{FTFQ}).
From this figure we clearly see that some non trivial cancellations 
occur and the high momentum components of the form factors
play an important role in  the whole range of $l$ considered.

Now we are ready to compare theoretical calculations for (\ref{Gamma})
 to  the four curves
corresponding to  different ``scenarios'' for the electric
form-factor behavior at large $Q^2$.
First of all, 
one finds that the overall magnitude and qualitative behavior with $l$
 are similar. 
This is very non-trivial,
since there are compensations between electric versus magnetic, and
small versus
large $Q^2$. 

Furthermore, our results seem to be able to distinguish
between various ``scenarios''.
In fact, the scenario 0 (in which $G_E\,\mu/G_M = 1$, identically)
and the scenario 3 (in which $G_E\,\mu/G_M $ rapidly approaches zero)
seem to be  ruled-out, assuming the theory prediction is correct.
Best agreement seem to be reached for the scenario 1,
the case in which $G_E\,\mu/G_M$ falls with $Q^2$ till about
 $0.6$, as found by Jlab experiment, and then
more or less stabilizes at large momenta.
Scenarios in which 
the ratio stabilizes at similar values are of course also  
possible.

\section{Discussion and comparison with Hadronic Models}
\label{di-quark}

After having calculated of the relevant Green functions, analytically
in SIA and numerically for the ensembles,  and compared our prediction
with the data,
let us go into somewhat more general discussion of the physics involved.

First let us ask, why  RILM and and IILM give so close
results for the ratio $\Gamma(l)$, but not the correlators themselves?
The former model has no correlation between instantons while the
latter has strong ones\footnote{To remind the reader about importance 
of
these correlation: in IILM  topological charge is screened and
topological
susceptibility is small, $O(m_q)$, as in the QCD vacuum with light
quarks. In RILM it is large, like in gluodynamics (quenched
QCD).} and their predictions are usually not that close.
The reason is that (see also discussion in section \ref{larged}), 
at the scale we are working at, the correlation
functions are dominated by \emph{one single instanton}.

This implies two important consequences. First, there actually exists a ``working window''
where the (unknown) continuum contribution is highly suppressed and yet SIA
is still accurate. Second, since the only relevant dimensionful parameter
in SIA is the instanton size $\rho$ (the instanton density also drops out in
the ratio), it follows that the scale of the nucleon form-factor
is completely determined by it\footnote{Note that
this feature is similar to what has been found 
for the pion form-factor in \cite{blotz}.}.  

The  content of the typical nucleon configuration is pictured
phenomenologically in different ways: e.g. the usual quark model
picture is that of three massive ``constituent'' quarks, bound by
confining forces. Another possibility discussed in literature is  
as a loose bound state of a quark with a scalar system
of two quarks, more tightly bound  (di-quark). 
In this section, we consider the problem of qualitatively 
determining the size of  such di-quark, in the instanton-based models.

From the Pauli principle on the level of instanton zero modes, only
two (u and d) 
 of the three quarks can be in the state described by the zero-mode 
wave-function (\ref{zmodes}) and therefore be localized around the
closest instanton. 
As a result, the typical size of the di-quark should 
 be determined by the  size of the instanton,
and be generally comparable to it.
 
 There is an interesting theoretical argument suggested in
\cite{Rapp:1998zu}, which relates pions and scalar di-quarks
in the QCD with two colors. Indeed, in this theory
di-quarks are not much heavier than mesons, but are
 degenerate with them, due to the so called Pauli-Gursey symmetry.
In particular, the lowest di-quarks should be  bound as
 strongly
as  the lowest meson, the massless pions. 
And indeed, the general pattern of symmetry breaking  
$SU(2N_f)\rightarrow Sp(2N_f)$  and the number of Goldstone modes is 
$$N_{goldstones}=2 N_f^2 - N_f -1.$$
 For the most
interesting case  $N_f=2$ there are five massless modes: 
three  pions, plus scalar di-quark $S$ and its anti-particle $\bar S$.
So, in this theory the pions and scalar di-quarks are related by
symmetry,
and should be truly identical. For three colors it is of course not so,
but by continuity 
in the number of colors, one may expect some traces of that to remain valid.

To get  quantitative estimates,  we proceed as follows.
On the one hand, we evaluate from the ILM the di-quark two-point function, 
\be
\Pi_{dq}(l) = <0|\,J^a_{sc}(x)\,J^\dagger_{a\,sc}(0)\,|0>,
\ee
where $J_{sc}^a(x)$ is the scalar current:
\be
J_{sc}^a (x)= \epsilon^{a\,b\,c}\, u^b(x)\,(\,C\,\gamma_5)\,d^c(x), 
\ee
and the di-quark three-point function,
\be
\Pi_{dq\,4}(l,l)=<0|J^a_{sc}(x/2)\,J^{em}_4(y)\,J^\dagger_{a\,sc}(-x/2)|0>,
\ee
and obtain the ratio $\Gamma_{dq}(l):=\Pi_{4\,dq}(l,l)/\Pi_{dq}(l)$.

\begin{figure}
\vskip 0.2in
\includegraphics[width=2.5 inc]{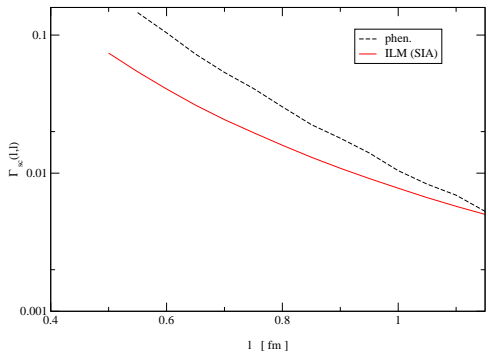}
\vskip 0.2in
\caption[]{
 \label{sdq3pfig}
The ratio $\Gamma_{sc\,4}(l,l)$ evaluated in the 
ILM, SIA (solid line) and from the ansatz (\ref{monopoledq}) with $\alpha=1760 MeV$ (dashed line).}
 \end{figure}

On the other hand, we compute the spectral decomposition for the
same quantities, as if the di-quark
 be  a physical (scalar) particle.
This assumption was discussed in detail in \cite{baryon2pt}, where the mass of 
the di-quark  was also determined from the RILM to be 
$M_{dq}=420 \pm 30\,\, MeV$.
As in the case of the nucleon, the large distance limit of
the spectral decomposition of the 
three-point function requires the knowledge of the form-factor 
of lowest lying state.
Exploiting the similarities between the physical properties of the pion 
and those of the di-quark in the ILM, we take a mono-pole shape ansatz for 
its form-factor,
\be
\label{monopoledq}
F_{dq}(Q^2)=\frac{1}{1+\frac{Q^2}{\alpha}}.
\ee
The free parameter $\alpha$ can be determined by 
comparing the function $\Gamma_{dq}$ evaluated 
(at large distances) from ILM and form the spectral decomposition
(see fig. \ref{sdq3pfig})
we found $\alpha\sim (1700 MeV)^2$.

Clearly, the agreement of ILM prediction with the pole contribution
to the spectral representation 
(obtained using (\ref{monopoledq}))
is definitely worse for the scalar
di-quark 
than for the pion \cite{blotz}, indicating that the simple 
mono-pole ansatz (\ref{monopoledq})
does not quite work in the case of the di-quark.
Of course this should not surprise:
after all, the di-quark is not an asymptotic state and we expect its charge 
distribution to be somewhat distorted by the presence of the third quark.

Nevertheless, from this fit we can still obtain 
a qualitative estimate of its size from the usual relation:
\be
\label{dqradius}
<r^2> := \, 6\,\left(\,\frac{d F_{dq}(q^2)}{d\,q^2}\, \right) = 
\frac{6}{\alpha^2} \simeq 0.3 fm. 
\ee
This value can, for instance, be compared with the 
phenomenological quark-diquark model  \cite{di-quark2},
 where  the value $<r_{dq}^2>\simeq \,(0.45 \,fm)^2$ was found.

Although both values can only be considered as qualitative estimates, they
 seem to indicate that the size of the di-quark is of the same  order
 of the typical size of an instanton.


\section{Conclusions and Outlook}
\label{conclusions}

In this paper we have studied the instanton contribution to the 
proton's electro-magnetic form-factors. 
We have  calculated the ratio $\Gamma_{sc}(l)$ of the proton electro-magnetic
three-point function (\ref{threepoint})
and the two-point function (\ref{twopoint}) in the RILM, in the IILM and, 
analytically, in the SIA.
It is important to stress that there is
no free parameters  and therefore the results of fig. \ref{Gammafig} 
represent an absolute prediction for this Green function. Furthermore,
the results happen to be independent on the details of the model,
and (like it was previously found for the pion form-factor)
only the instanton mean radius $\rho\sim 1/600 \,Mev$  is actually 
setting the scale of it.

We have compared our theoretical predictions with
the phenomenological curve obtained from the nucleon's pole 
contribution to spectral representation of the correlation functions.
In particular, in order to evaluate the proton contribution to the 
electro-magnetic three-point function we have used several fits of the 
$G_E\,\mu/G_M$  data.
These fits describe scenarios in which such ratio reaches 
different asymptotic values, in the kinematical region under current 
investigation at JLAB.
We found that our prediction, is consistent with the deviation
 from the dipole shape of $G_E$ 
and is very much in agreement with a scenario in which the $G_E\,\mu /G_M$ 
ratio stabilizes somewhere around $0.6$, at high $Q^2$.

We have also performed a similar analysis to obtain
 a qualitative  estimate of the size of the di-quark part of the nucleon.  
We found $<r^2>\sim (0.3\,fm)^2$, which is 
consistent with the picture in which 
quark-zero modes are localized around the closest 
instanton and in reasonable 
agreement with the value $<r^2>\sim (0.45\,fm)^2$ found in \cite{di-quark2}.

Our analysis revealed that, although the RILM and the IILM give in general
 quite different correlators, their 
prediction for the ratio $\Gamma (l)$ is exactly the same.
This implies that at the moment of scattering two out of three
quarks interact essentially with one instanton, via its zero mode.
This particular configuration is, in a way, at a core of
quark-diquark model.

As we found that it is quite possible to perform 
an analysis of the  pion and proton form-factor in  the simplest SIA,
without needing to model the continuum contribution,
one can hope that such analytical results can be Fourier transformed to
momentum space and allow for direct comparison with experimental data.
We intend to exploit this possibility in our further works.

\acknowledgments
We would like to thank T.Sch\"afer for several discussions and numerical help.
The work is partly supported by the US DOE grant No. DE-FG02-88ER40388.

\newpage
\appendix
\section{Derivation of the nucleon pole contribution to the electro-magnetic 
three-point function}
\label{deriv}

We want to evaluate the proton's pole contribution to the three-point function:
\begin{equation}	
\Pi_{sc \, 4}(x,y)=<0|\, Tr\,[ \, \eta_{sc}(x) \, J^{em}_4(y)\bar{\eta}_{sc}(0) \, \gamma_4 \,]\,|0>,
\end{equation}
where $J^{em}_4$ is the electro-magnetic current:
\begin{equation}
J^{em}_4(y):= \frac{2}{3}\, \bar{u}(y)\,\gamma_4\, u(y) -  \frac{1}{3}\,
 \bar{d}(y)\,\gamma_4 d(y),
\end{equation}.
and we haven chosen $x_4\ge y_4\ge 0$.

We begin by inserting two complete sets of states:
\be
\Pi_{sc \, 4}(x,y)= 
\sum_{\Gamma '}\, \sum_\Gamma 
Tr \left[ <0| \,\eta_{sc}\left(\frac{x}{2}\right)| 
\Gamma '>
\right.\nonumber\\
\left.<\Gamma '|J_4(y)|\Gamma ><\Gamma | \bar{\eta}_{sc}
\left(-\frac{x}{2}\right)|0>
\gamma_4\right].
\ee
The term of this double sum that depends  only on the lowest lying 
state (nucleon) represents
the pole, while the sum of all other terms forms the continuum.
We therefore obtain:
\be
\Pi_{sc\, 4}(x,y) =
\sum_{s, s'} 
\int \frac{d^3\, p'}{2\,\omega_{p'} \, (2\pi)^3}
\int \frac{d^3\, p}{2\,\omega'_{p} \, (2\pi)^3}
\nonumber\\
\cdot\,Tr \,[\,<0| \,\eta_{sc}\,\left(\,\frac{x}{2}\,\right)\,|\, p',s' ;M_N\,>
\cdot\nonumber\\
\cdot\,<\,p',s';M_N\,|\,J_4(y)\,|\,p,s;M_N\,>\cdot
\nonumber\\
\cdot\,<\,p,s;M_N\,|\, \bar{\eta}_{sc}\left(-\frac{x}{2}\right)\,|\,0\,>
\,\gamma_4\, ]+... \,,
\ee
where 
$M_N$ denotes the proton's rest mass and
the ellipses denote the continuum contribution.
Now, using equation (\ref{lambdas}) and recalling the 
definition of Pauli and Dirac form-factors,
\be
<p',s';M_N|J_\mu (0) |p,s ;M_N>\,=
\nonumber\\
= \bar{u}(p',s')  
\,[\gamma_\mu F_1(Q^2) +
\frac{i\, M_N}{2} \sigma_{\mu\,\nu} q^\nu\,
F_2(Q^2)] \,u(p,s),
\ee
we get:
\be
\Pi_{sc,\,4}(x,y)=
\int \frac{d^3\, p'}{2\,\omega_{p} \, (2\pi)^3}
\int \frac{d^3\, p}{2\,\omega'_{p'} \, (2\pi)^3}
\,\Lambda_{sc}^2\cdot
\nonumber\\
\,exp
\left[-i\,\frac{x}{2}\,(p + p') - i\, y\,(p-p')
\right]\,      
Tr \left[ \,( \ds{p}'+ M_N )\,\cdot
\right.
\nonumber\\
\left.
\left( \gamma_4 \,F_1(Q^2) +
\frac{i M_N}{2}\sigma_{4\,\nu} q^\nu
F_2(Q^2)\right)
(\ds{p} + M_N )\gamma_4 \right]+...
\ee 
Now we can re-write the form-factors in terms of their Fourier transform,
\be
\label{FTFQ}
F_1(Q^2) &=& \int\,\frac{d^4 z}{(2\pi)^4}\, e^{i\,q \cdot z}\, F_1(z)\\
F_2(Q^2) &=& \int\,\frac{d^4 z}{(2\pi)^4}\, e^{i\,q \cdot z}\, F_2(z),
\ee
and introduce the expression for the nucleon propagator in coordinate space
( $ x'_0\ge x_0$ ):
\be
S_{M_N}(x',x) =\int\,\frac{d^3 p}{(2\pi)^3 \, 2 \, \omega_p}\, e^{-i\,p \cdot (x'-x)}
( \ds{p} + M_N ),
\ee
obtaining:
\be
\label{res}
\Pi_{sc\,4} (x,y) = \Lambda_{sc}^2 \int\,d^4 z\, Tr\,
\left[ 
S_{M_N}\left(\frac{x}{2}, y+z \, \right)\cdot\right.\nonumber\\
\left.\left[\gamma_4 F_1(z)
+ \frac{i\sigma_{4\,\nu}} {2\,M_N} F_2^\nu (z)\right] 
S_{M_N}\left(y+z,-\frac{x}{2}  \right)\gamma_4
\right] + ...\nonumber 
\\
\ee
where:
\be
F_2^\nu(z) := i\,\partial^\nu\, F_2(z).
\ee

One can check that the analytic continuation of
(\ref{res}) to Euclidean space
\footnote{ In our Euclidean formulation, 
$x\cdot y= \sum_{\mu=1}^4 x_\mu\,y_\mu$, 
$\{\,\gamma_\mu,\gamma_\nu\,\} = 2\,\delta_{\mu\,\nu}$,
$\gamma_\nu^\dagger= \gamma_\nu$ and 
$\sigma_{\mu\,\nu}= \frac{i}{2}[\,\gamma_\mu , \gamma_\nu\,]$.} 
is real. 


\end{document}